\documentclass{appolb}
\usepackage{graphicx}
\begin{document}
\title{Quantum phase transition in a transverse Ising chain\\
       with regularly varying parameters
\thanks{Presented at the Strongly Correlated Electron Systems
        Conference, Krak\'ow 2002}%
}

% Authors and Affiliations

\author{
O.Derzhko$^{(1)}$, J.Richter$^{(2)}$, T.Krokhmalskii$^{(1)}$, O.Zaburannyi$^{(1)}$
\address{$^{(1)}$ Institute for Condensed Matter Physics,\\
                  1 Svientsitskii Street, L'viv-11, 79011, Ukraine\\
         $^{(2)}$ Institut f\"{u}r Theoretische Physik,
                  Universit\"{a}t Magdeburg,\\
                  P.O. Box 4120, D-39016 Magdeburg, Germany}
       }
\maketitle

% Abstract

\begin{abstract}

Using rigorous analytical analysis and  exact numerical data
for the spin-$\frac{1}{2}$ transverse Ising chain
we discuss the effects of regular alternation
of the Hamiltonian parameters
on the quantum phase transition inherent in the model.

\end{abstract}


% The main text

The one-dimensional spin-$\frac{1}{2}$ Ising model in a transverse field
(the transverse Ising chain)
defined by the Hamiltonian
\begin{eqnarray}
\label{01}
H=\sum_n2Is_n^xs_n^x+\sum_n\Omega s_n^z
\end{eqnarray}
is known to be the simplest system
exhibiting a quantum (zero-temperature) phase transition
driven by the transverse field \cite{cit1}.
Most of the performed studies for this model
use the exact eigenvalues and eigenfunctions
of its Hamiltonian (\ref{01})
that makes the problem amenable for rigorous analysis \cite{cit1,cit2}.
It is generally known
that the critical value of the transverse field is
$\Omega_c=\vert I\vert$
(and $\Omega_c=-\vert I\vert$).
The longitudinal (Ising) magnetization per site
$m^x=\frac{1}{N}\sum_n\langle s_n^x\rangle$
is the order parameter of the system.
$\vert m^x\vert$ varies
from $\frac{1}{2}$ (for $\Omega=0$)
to 0 (for $\Omega\ge\Omega_c$)
according to 
$\vert m^x\vert 
=\frac{1}{2}\left(1-\left(\frac{\Omega}{\Omega_c}\right)^2\right)^{\frac{1}{8}}$.
The quantum phase transition at $\Omega_c$
is equivalent
to the thermal phase transition
of the square-lattice Ising model.
After the understanding of the properties of the basic model 
was achieved
the models with various modifications were introduced
and the effects of introduced changes
on the quantum phase transition were discussed.
Among numerous works in this field one may mention
an analysis of the critical behaviour 
of the aperiodic transverse Ising chain 
by J.M.Luck \cite{cit3},
an extensive real-space renormalization-group treatment
of the random transverse Ising chain
by D.S.Fisher \cite{cit4}
or
renormalization-group study
of the aperiodic transverse Ising chain
by F.Igl\'{o}i et al \cite{cit5}.
It should be remarked, however,
that the study of a simpler case
of the {\em regularly inhomogeneous} transverse Ising chain
is still lacking
although many properties of such a system
can be examined exploiting fermionic representation
either rigorously analytically
(using continued fractions)
or on a very precise and well-controlled level of approximation numerically
(studying long chains).

In this brief report 
(for the extended version see \cite{cit6})
we highlight
how deviations from the pure uniform crystalline system
in kind of regularly alternating
exchange interactions and transverse fields
influence the quantum phase transition
inherent in the spin-$\frac{1}{2}$ transverse Ising chain.
The regular alternation is obtained 
by substituting in (\ref{01})
$I_n$ and $\Omega_n$
instead of
$I$ and $\Omega$
assuming a periodic sequence
$$I_1\Omega_1\ldots I_p\Omega_pI_1\Omega_1\ldots I_p\Omega_p\ldots .$$
In particular, 
we present the results  
of rigorous analytical study based on \cite{cit7}
for thermodynamic quantities 
and
of exact numerical study using the method illustrated in \cite{cit8}
for spin correlations.
Our main conclusions are
that the number of second-order quantum phase transitions
for a given period of alternation $p$ 
strongly depends on the concrete values of the Hamiltonian parameters
whereas the critical behaviour remains as for the uniform chain.
Moreover,
for a certain values of the Hamiltonian parameters 
weaker singularities of the ground-state quantities 
may appear. 

We start from recalling an old result of P.Pfeuty \cite{cit9}
in the present context.
P.Pfeuty showed
that for the spin-$\frac{1}{2}$ transverse Ising chain Hamiltonian
the gap in the excitation spectrum
(in the thermodynamic limit)
is zero
at the ``critical point''
\begin{eqnarray}
\label{02}
\prod_nI_n=\prod_n\Omega_n.
\end{eqnarray}
A result similar to (\ref{02})  
can be also derived by the continued fraction approach
developed in Ref. \cite{cit7}.
In particular, 
for a chain of period 2 (period 3) Eq. (\ref{02}) yields
$I_1I_2=\pm\Omega_1\Omega_2$
($I_1I_2I_3=\pm\Omega_1\Omega_2\Omega_3$).
Consider further chains with regularly alternating transverse field 
$\Omega_n=\Omega+\Delta\Omega_n$,
$\Delta\Omega_1+\ldots+\Delta\Omega_p=0$,
$\vert I_n\vert=\vert I\vert\;(=1)$.
For a chain of period 2 
($\Delta\Omega_1=-\Delta\Omega_2=\Delta\Omega>0$)
the equation 
$\Omega_c^2-{\Delta\Omega}^2=\pm I^2$
yields 
either two critical transverse fields,
$\pm\sqrt{{\Delta\Omega}^2+I^2}$
(if $\Delta\Omega<\vert I\vert$),
or three critical transverse fields,
$-\sqrt{2}\Delta\Omega$, $0$, $\sqrt{2}\Delta\Omega$
(if $\Delta\Omega=\vert I\vert$),
or 
four critical transverse fields,
$\pm\sqrt{{\Delta\Omega}^2+I^2}$ and $\pm\sqrt{{\Delta\Omega}^2-I^2}$
(if $\Delta\Omega>\vert I\vert$).
Thus,
the number of critical transverse fields 
yielding gapless energy spectrum 
of the Ising chain in a modulated transverse field of period 2 
depends on a strength of inhomogeneity.
For small $\Delta\Omega$ only quantitative changes 
with respect to the homogeneous case may be expected,
whereas for large $\Delta\Omega$ some qualitative changes should occur.
In Fig. 1 
\begin{figure}[!ht]
\begin{center}
\includegraphics[width=1.0\textwidth]{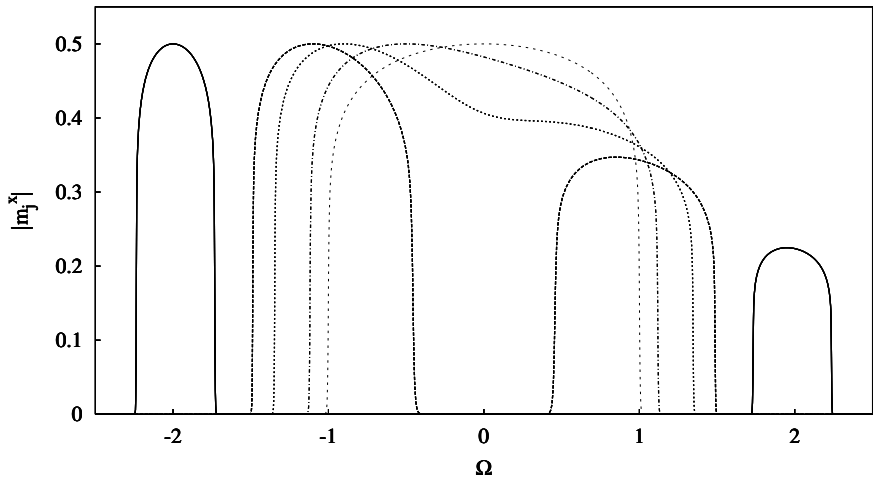}
\end{center}
\caption{
The ground-state longitudinal sublattice magnetization
of the spin-$\frac{1}{2}$ transverse Ising chain
with $\vert I_1\vert=\vert I_2\vert=1$,
$\Omega_{1,2}=\Omega\pm\Delta\Omega$
in dependence on the transverse field $\Omega$;
$\Delta\Omega=0$ (thin dashed curve),
$\Delta\Omega=0.5$ (dot-dashed curve),
$\Delta\Omega=0.9$ (dotted curve),
$\Delta\Omega=1.1$ (dashed curve),
$\Delta\Omega=2$ (solid curve).
The on-site magnetizations were obtained from the correlation functions
$\langle s^x_{100}s^x_{600}\rangle$
and
$\langle s^x_{100}s^x_{601}\rangle$
computed for chains which consist of $N=700$ sites.}
\label{fig1}
\end{figure}
we report a behaviour 
of the ground-state longitudinal magnetization 
at sites at which the transverse field equals to $\Omega+\Delta\Omega$
as a function of $\Omega$ 
for different values of 
$\Delta\Omega=0,\;0.5,\;0.9$ 
($\Delta\Omega<\vert I\vert$)
and 
$\Delta\Omega=1.1,\;2$ 
($\Delta\Omega>\vert I\vert$)
(for details of numerical calculation see \cite{cit8}).
The data show 
that for $\Delta\Omega<\vert I\vert$
two phases appear as $\Omega$ varies:
the quantum Ising phase 
for $\vert\Omega\vert<\sqrt{{\Delta\Omega}^2+I^2}$ 
and
the quantum paramagnetic phase otherwise   
(thin dashed, dot-dashed and dotted curves in Fig. \ref{fig1}).
Contrary,
for $\Delta\Omega>\vert I\vert$
three phases appear as $\Omega$ varies:
the low-field quantum paramagnetic phase 
for $\vert\Omega\vert<\sqrt{{\Delta\Omega}^2-I^2}$, 
the quantum Ising phase 
for 
$\sqrt{{\Delta\Omega}^2-I^2}
<\vert\Omega\vert
<\sqrt{{\Delta\Omega}^2+I^2}$ 
and
the strong-field quantum paramagnetic phase otherwise   
(dashed and solid curves in Fig. \ref{fig1}).
The quantum phase transitions between different phases 
are accompanied by divergence of the correlation length $\xi^x$
and logarithmic singularity of the static transverse susceptibility $\chi^z$
(these results follow from the numerical data 
and the analytical formula for the ground-state energy 
obtained with the help of continued fractions,
respectively \cite{cit6}).
Thus,
the critical behaviour remains as for the uniform chain.
Next we turn to the low-temperature behaviour of the specific heat $c$.
The existence of zero-energy excitations 
produces the linear dependence $c$ versus $T$ as $T\to 0$
which may serve as an indication of the quantum critical point.
The exact analytical results 
for the temperature dependence $c$ versus $T$ 
obtained with the help of continued fractions 
confirm 
that there are 
either 
two (if  $\Delta\Omega<\vert I\vert$)
or 
four (if $\Delta\Omega>\vert I\vert$)
values of transverse field 
at which $c$ decays linearly as $T\to 0$ 
\cite{cit6}.
If $\Delta\Omega=\vert I\vert$
besides the critical fields $\pm\sqrt{2}\Delta\Omega$ 
one more critical field, $\Omega_c=0$, appears.
While $\Omega$ approaches $\Omega_c=0$ 
the energy gap vanishes 
$\sim\vert\Omega-\Omega_c\vert^2$
(but not 
$\sim\vert\Omega-\Omega_c\vert$
as for the discussed before critical points $\Omega_c$)
that results in a finite value of 
$\chi^z\sim\vert\Omega-\Omega_c\vert^2\ln\vert\Omega-\Omega_c\vert$ 
and a logarithmic singularity of its second derivative
at $\Omega=\Omega_c=0$.

Extending the reported analysis for chains of period 3 
we find that either two, or four, or six 
second-order quantum phase transition points may occur.
Again the number of the quantum phase transitions 
is controlled by the strength of nonuniformity 
and the critical behaviour remains the same as for the uniform case.
Moreover,
weaker singularities may appear.
It is worth to note
that the condition (\ref{02}) may be tuned by arbitrary parameter(s) 
influencing on-site transverse fields and intersite exchange interactions.
For example,
for a chain with 
$I_{1,2}=I\pm\Delta I$, 
$\Delta I\ge 0$,
$\Omega_n=\Omega$
the change of $I$ 
may yield a different number of quantum phase transition points 
depending 
on a relation between $\Delta I$ and $\Omega$.
Finally,
let us mention recent papers \cite{cit10,cit11}
where similar questions have been addressed 
for spin-$\frac{1}{2}$ anisotropic $XY$ chains.

The present study was partly supported by the DFG
(project 436 UKR 17/1/02).
O. D. acknowledges the kind hospitality of the Magdeburg University
in the autumn of 2002.
O. D. and T. K. were supported by the STCU
under the project \#1673.

\end{document}